\documentclass[10pt]{article}

\usepackage{amsmath}
\usepackage{amssymb}

\usepackage{graphicx}

\usepackage{cite}

\usepackage{color} 

\usepackage{setspace} 
\graphicspath{{./}{./fig/}}

\usepackage[labelfont=bf,labelsep=period,justification=raggedright]{caption}

\makeatletter
\renewcommand{\@biblabel}[1]{\quad#1.}
\makeatother

\date{}

\begin{document}

\begin{flushleft}
{\Large
\textbf{Potentials of mean force for protein structure prediction
vindicated, formalized and generalized}
}
\\
Thomas Hamelryck$^{1,\dagger,\#}$, 
Mikael Borg$^{1,\#}$, 
Martin Paluszewski$^{1,\#}$,
Jonas Paulsen$^{1}$, 
Jes Frellsen$^{1}$, 
Christian Andreetta$^{1}$,
Wouter Boomsma$^{2,3}$, 
Sandro Bottaro$^{2}$,
Jesper Ferkinghoff-Borg$^{2,\ddagger}$
\\
\bf{1} Bioinformatics Center, Department of Biology, University of
Copenhagen, Copenhagen, Denmark
\\
\bf{2} Biomedical engineering, DTU Elektro, Technical University
of Denmark, Lyngby, Denmark
\\
\bf{3} Department of Chemistry, University of Cambridge, Cambridge, United
Kingdom
\\
$\dagger$ E-mail: thamelry@binf.ku.dk \newline
$\ddagger$ E-mail: jfb@elektro.dtu.dk
\\
$\#$ Joint first authors.
\end{flushleft}

\section*{Abstract}
Understanding protein structure is of crucial importance in science,
medicine and biotechnology. For about two decades, knowledge based
potentials based on pairwise distances -- so-called {}``potentials of
mean force'' (PMFs) -- have been  center stage in the prediction and
design of protein structure and the simulation of protein folding.
However, the validity, scope and limitations of these potentials are
still vigorously debated and disputed, and the optimal choice of the
reference state -- a necessary component of these potentials -- is an
unsolved problem.  PMFs are loosely justified by analogy to the
reversible work theorem in statistical physics, or by a statistical
argument based on a likelihood function. Both justifications are
insightful but leave many questions unanswered. Here, we show for the
first time that PMFs can be seen as approximations to quantities that do
have a rigorous probabilistic justification: they naturally arise when
probability distributions over different features of proteins need to be
combined. We call these quantities "reference ratio distributions"
deriving from the application of the "reference ratio method".
This new view is not only of theoretical relevance, but leads to many
insights that are of direct practical use: the reference state is
uniquely defined and does not require external physical insights; the
approach can be generalized beyond pairwise distances to arbitrary
features of protein structure; and it becomes clear for which purposes
the use of these quantities is justified. We illustrate these insights
with two applications, involving the radius of gyration and hydrogen
bonding. In the latter case, we also show how the reference ratio
method can be iteratively applied to sculpt an energy funnel. Our
results considerably increase the understanding and scope of energy
functions derived from known biomolecular structures.


\section*{Introduction}

Methods for protein structure prediction, simulation and design rely
on an energy function that represents the protein's free energy landscape;
a protein's native state typically corresponds to the state with minimum
free energy \cite{pmid4124164}. So-called knowledge based potentials
(KBP) are parametrized functions for free energy calculations that
are commonly used for modeling protein structures \cite{moult1997comparison,pmid17075131}.
These potentials are obtained from databases of known protein structures
and lie at the heart of some of the best protein structure prediction
methods. The use of KBPs originates from the work of Tanaka and Scheraga
\cite{pmid1004017} who were the first to extract effective interactions
from the frequency of contacts in X-ray structures of native proteins.
Miyazawa and Jernigan formalized the theory for contact interactions
by means of the quasi-chemical approximation \cite{miyazawa1985estimation,miyazawa1999empirical}.

Many different approaches for developing KBPs exist, but the most
successful methods to date build upon a seminal paper by Sippl --
published two decades ago -- which introduced KBPs based on probability
distributions of pairwise distances in proteins and reference states
\cite{pmid2359125}. These KBPs were called {}``potentials of mean
force'', and seen as approximations of free energy functions. Sippl's
work was inspired by the statistical physics of liquids, where a {}``potential
of mean force'' has a very precise and undisputed definition and
meaning \cite{chandler1987,mcquarrie2000}. However, the validity
of the application to biological macromolecules is vigorously disputed
in the literature
\cite{finkelstein1995protein,rooman1995database,pmid8609636,moult1997comparison,ben-naim:3698,pmid9526121,shortle2003propensities,kirtay2005knowledge,muegge2006pmf}.
Nonetheless, PMFs are widely used with considerable success; not only
for protein structure prediction \cite{pmid9149153,pmid17075131,colubri2006minimalist},
but also for quality assessment and identification of errors \cite{pmid8108378,eramian2006composite,rykunov2010new},
fold recognition and threading \cite{pmid1614539,pmid19291741}, molecular
dynamics \cite{pmid19291741}, protein-ligand interactions
\cite{gohlke2000knowledge,kirtay2005knowledge},
protein design and engineering \cite{gilis1997predicting,gilis2000popmusic},
and the prediction of binding affinity \cite{muegge2006pmf,su2009quantitative}.
In this article, the abbreviation {}``PMF'' will refer to the pairwise distance
dependent KBPs following Sippl \cite{pmid2359125}, and the generalization
that we introduce in this article; we will write {}``potentials of mean
force'' in full when we refer to the real, physically valid potentials
as used in liquid systems
\cite{ben-naim:3698,mcquarrie2000,pmid16193038}. At the end of the
article, we will propose a new name for these statistical quantities, to
set them apart from true potentials of mean force with a firm physical basis. 

Despite the progress in methodology and theory, and the dramatic
increase in the number of experimentally determined protein structures,
the accuracy of the energy functions still remains the main obstacle to
accurate protein structure prediction
\cite{pmid18767152,shmygelska2009generalized,rykunov2010new}.  Recently,
several groups demonstrated that it is the quality of the coarse grained
energy functions \cite{pmid9149153}, rather than inadequate
sampling, that impairs the successful prediction of the native state
\cite{pmid18767152,shmygelska2009generalized}. The insights presented in
this article point towards a new, theoretically well-founded way to
construct and refine energy functions, and thus address a timely
problem.

We start with an informal outline of the general ideas presented in this
article, and then analyze two notable attempts in the literature to
justify PMFs. We point out their shortcomings, and subsequently present
a rigorous probabilistic explanation of the strengths and shortcomings
of traditional pairwise distance PMFs. This explanation sheds a
surprising new light on the nature of the reference state, and allows
the generalization of PMFs beyond pairwise distances in a statistically
valid way. Finally, we demonstrate our method in two applications
involving protein compactness and hydrogen bonding. In the latter case,
we also show that PMFs can be iteratively optimized, thereby effectively
sculpting an energy funnel
\cite{bryngelson1987spin,leopold1992protein,dill1997levinthal,pmid12926006,fain2003funnel,pmid19291741}.

\section*{Results and discussion}

\subsection*{Overview}
\begin{figure}[!ht]
\begin{centering}
\includegraphics[clip,width=0.9\columnwidth]{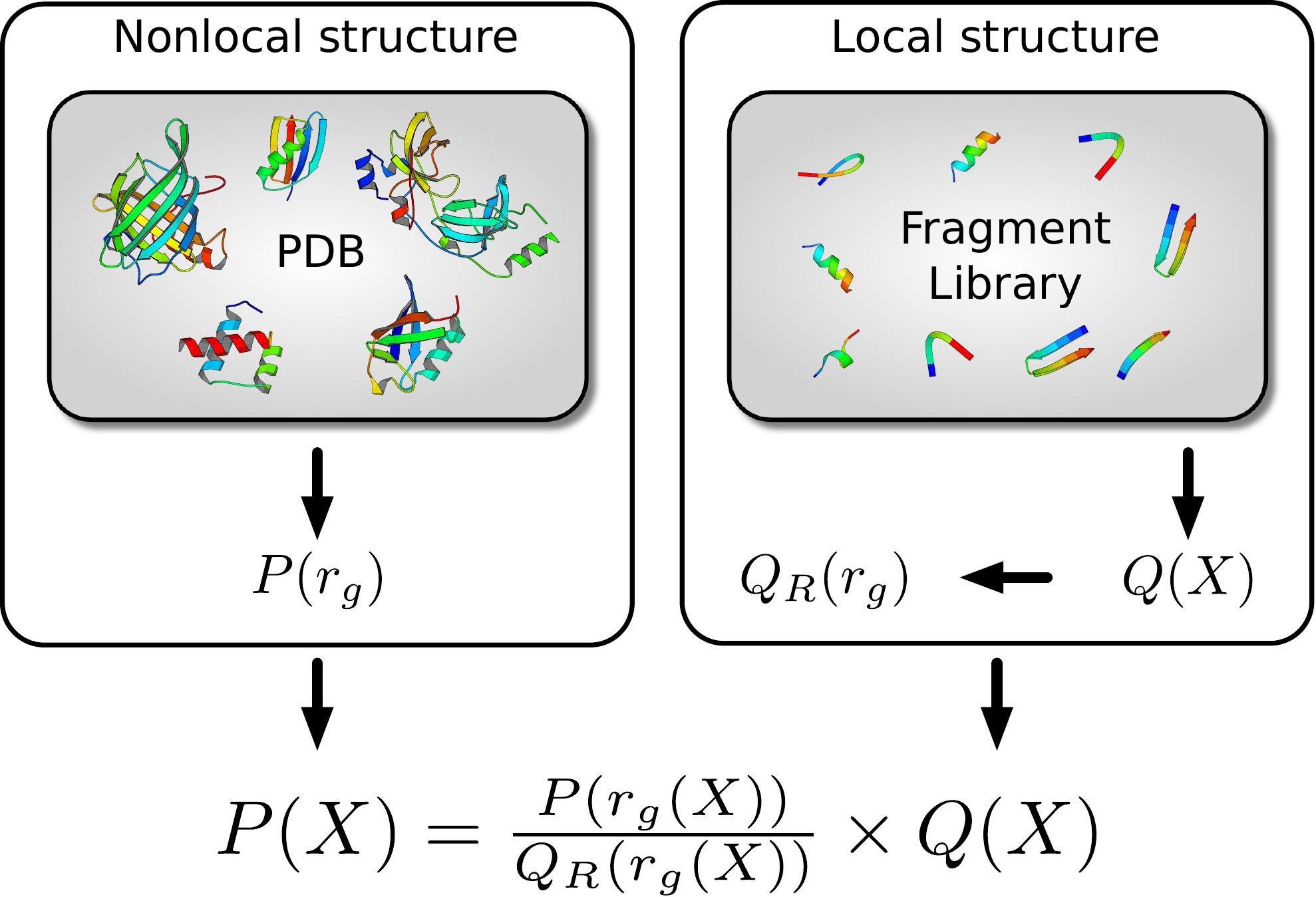} 
\par\end{centering}

\caption{\label{fig:simple}\textbf{Illustration of the central idea
presented in this article.} In this example, the goal is to sample
conformations with a given distribution $P(r_{g})$ for the radius of
gyration $r_{g}$, and a plausible local structure.  $P(r_{g})$ could,
for example, be derived from known structures in the Protein Data Bank
(PDB, left box). $Q(X)$ is a probability distribution over local
structure $X$, typically embodied in fragment library (right box). In
order to combine $Q(X)$ and $P(r_{g})$ in a meaningful way (see text),
the two distributions are multiplied and divided by $Q_{R}(r_{g})$
(formula at the bottom); $Q_{R}(r_{g})$ is the probability distribution
over the radius of gyration for conformations sampled solely from the
fragment library (that is,  $Q(X)$).  The probability distribution
$P(X)$ will generate conformations with plausible local structures (due
to $Q(X)$), while their radii of gyration will be distributed according
to $P(r_{g})$, as desired. This simple idea lies at the theoretical
heart of the PMF expressions used in protein structure prediction.}  

\end{figure}

In order to emphasize the practical implications of the theoretical
insights that we present here, we start with a very concrete example
that illustrates the essential concepts (see Fig.\ \ref{fig:simple}).
Currently, protein structure prediction methods often make use of
fragment libraries: collections of short fragments derived from known
protein structures in the Protein Data Bank (PDB). By assembling a
suitable set of fragments, one obtains conformations that are
protein-like on a local length scale.  That is, these conformations
typically lack non-local features that characterize real proteins, such
as a well-packed hydrophobic core or an extensive hydrogen bond network.
Such aspects of protein structure are not, or only partly, captured by
fragment libraries.  

Formally, a fragment library specifies a
probability distribution $Q(X)$, where $X$ is for example a vector of
dihedral angles. In order to obtain conformations that also possess the
desired non-local features, $Q(X)$ needs to be complemented with another
probability distribution $P(Y)$, with $Y$ being for example a vector of
pairwise distances, the radius of gyration, the hydrogen bonding
network, or any combination of non-local features. Typically, $Y$ is a
deterministic function of $X$; we use the notation $Y(X)$ when necessary. 

For the sake of argument, we will focus on the radius of gyration ($r_g$) at
this point; in this case $Y(X)$ becomes $r_g(X)$. We assume that a suitable
$P(r_{g})$ was derived from the set of known protein structures; without loss
of generality, we leave out the dependency on the amino acid sequence for
simplicity. The problem that we address in this article can be illustrated with
the following question: how can we combine $P(r_{g})$ and $Q(X)$ in a rigorous,
meaningful way? In other words, we want to use the fragment library to sample
conformations whose radii of gyration $r_{g}$ are distributed according to
$P(r_{g})$. These conformations should display a realistic \emph{local}
structure as well, reflecting the use of the fragment library.  
Simply multiplying $P(r_{g}(X))$ and $Q(X)$ does not lead to the
desired result, as $X$ and $R_{g}$ are not independent; the resulting
conformations will not be distributed according to $P(r_g)$.

The solution is given in Fig.\ \ref{fig:simple}; it involves the probability
distribution $Q_{R}(r_{g})$, the probability distribution over the radius of
gyration for conformations sampled solely from the fragment library. The
subscript $R$ stands for \emph{reference state} as will be explained below. The
solution generates conformations whose radii of gyration are distributed
according to $P(r_{g})$. The influence of $Q(X)$ is apparent in the fact that
for conformations with a given $r_{g}$, their local structure $X$ will be
distributed according to $Q(X\mid r_{g})$. The latter distribution has a clear
interpretation: it corresponds to sampling an infinite amount of conformations
from a fragment library, and retaining only those with the desired $r_g$. Note
that even if we chose the uniform distribution for $Q(X)$, the resulting
$Q_{R}(r_{g})$ will \emph{not} (necessarily) be uniform.  

Intuitively, $P(r_g)$ provides correct information about the radius of
gyration, but no information about local structure; $Q(X)$ provides
approximately correct information about the structure of proteins on a
local length scale, but is incorrect on a global scale (leading to an
incorrect probability distribution for the radius of gyration); finally,
the formula shown in Fig.\ \ref{fig:simple} merges these two
complementary sources of information together. Another viewpoint is that
$P(r_g)$ and $Q(r_g)$ are used to correct the shortcomings of $Q(X)$.
This construction is statistically rigorous, provided that $P(r_{g})$
and $Q(X)$ are proper probability distributions. 

After this illustrative example, we now review the use of PMFs in
protein structure prediction, and discuss how PMFs can be understood and
generalized in the theoretical framework that we briefly outlined here.

\subsection*{Pairwise PMFs for protein structure prediction}

Many textbooks present PMFs as a simple consequence of the Boltzmann
distribution, as applied to pairwise distances between amino acids.
This distribution, applied to a specific pair of amino acids,
is given by: \[
P\left(r\right)=\frac{1}{Z}e^{-\frac{F\left(r\right)}{kT}}\]
where $r$ is the distance, $k$ is Boltzmann's constant, $T$ is
the temperature and $Z$ is the partition function, with $Z=\int e^{-\frac{F(r)}{kT}}dr$.
The quantity $F(r)$ is the free energy assigned to the pairwise system.
Simple rearrangement results in the \emph{inverse Boltzmann formula},
which expresses the free energy $F(r)$ as a function of $P(r)$:

\[
F\left(r\right)=-kT\ln P\left(r\right)-kT\ln Z\]

\noindent To construct a PMF, one then introduces a so-called \emph{reference
state} with a corresponding distribution $Q_{R}$ and partition function
$Z_{R}$, and calculates the following free energy difference: \begin{equation}
\Delta F\left(r\right)=-kT\ln\frac{P\left(r\right)}{Q_{R}\left(r\right)}-kT\ln\frac{Z}{Z_{R}}\label{eq:classic}\end{equation}

\noindent The reference state typically results from a hypothetical
system in which the specific interactions between the amino acids
are absent \cite{pmid2359125}. The second term involving $Z$ and
$Z_{R}$ can be ignored, as it is a constant.

In practice, $P(r)$ is estimated from the database of known protein
structures, while $Q_{R}(r)$ typically results from calculations
or simulations. For example, $P(r)$ could be the conditional probability
of finding the $C\beta$ atoms of a valine and a serine at a given
distance $r$ from each other, giving rise to the free energy difference
$\Delta F$. The total free energy difference of a protein,
$\Delta F_{\textrm{TOT}}$, is then claimed to be the sum
of all the pairwise free energies:\begin{eqnarray}
\Delta F_{\textrm{TOT}} & = & \sum_{i<j}\Delta F(r_{ij}\mid a_{i},a_{j})\label{eq:Ftot}\\
 & = & -kT\sum_{i<j}\ln\frac{P\left(r_{ij}\mid a_{i},a_{j}\right)}{Q_{R}\left(r_{ij}\mid a_{i},a_{j}\right)}\end{eqnarray}

\noindent where the sum runs over all amino acid pairs $a_{i},a_{j}$
(with $i<j$) and $r_{ij}$ is their corresponding distance. It should
be noted that in many studies $Q_{R}$ does not depend on the amino
acid sequence \cite{rooman1995database}.

Intuitively, it is clear that a low free energy difference indicates
that the set of distances in a structure is more likely in proteins than
in the reference state. However, the physical meaning of these PMFs have
been widely disputed since their introduction
\cite{pmid8609636,moult1997comparison,ben-naim:3698,pmid9526121,shortle2003propensities}.
Indeed, why is it at all necessary to subtract a reference state energy?
What is the optimal reference state? Can PMFs be generalized and
justified beyond pairwise distances, and if so, how? Before we discuss
and clarify these issues, we discuss two qualitative justifications that
were previously reported in the literature: the first based on a
physical analogy, and the second using a statistical argument.

\subsection*{PMFs from the reversible work theorem}

The first, qualitative justification of PMFs is due to Sippl, and
based on an analogy with the statistical physics of liquids \cite{pmid9079391}.
For liquids \cite{chandler1987,pmid9079391,ben-naim:3698,pmid9526121,mcquarrie2000},
the potential of mean force is related to the \emph{pair correlation
function} $g(r)$, which is given by: \[
g(r)=\frac{P(r)}{Q_{R}(r)}\]
where $P(r)$ and $Q_{R}(r)$ are the respective probabilities of
finding two particles at a distance $r$ from each other in the liquid
and in the reference state\emph{.} For liquids, the reference state
is clearly defined; it corresponds to the ideal gas, consisting of
non-interacting particles. The two-particle potential of mean force
$W(r)$ is related to $g(r)$ by: \begin{equation}
W(r)=-kT\log g(r)=-kT\log\frac{P(r)}{Q_{R}(r)}\label{eq:sippl}\end{equation}
According to the \emph{reversible work theorem}, the two-particle
potential of mean force $W(r)$ is the reversible work required to
bring two particles in the liquid from infinite separation to a distance
$r$ from each other \cite{chandler1987,mcquarrie2000}.

Sippl justified the use of PMFs -- a few years after he introduced
them for use in protein structure prediction \cite{pmid2359125} -- by
appealing to the analogy with the reversible work theorem for liquids
\cite{pmid9079391}. For liquids, $g(r)$ can be experimentally measured
using small angle X-ray scattering; for proteins, $P(r)$ is obtained
from the set of known protein structures, as explained in the previous
section. The analogy described above might provide some physical insight,
but, as Ben-Naim writes in a seminal publication \cite{ben-naim:3698}:
{}``the quantities, referred to as `statistical potentials,' `structure
based potentials,' or `pair potentials of mean force', as derived from
the protein data bank, are neither `potentials' nor `potentials of
mean force,' in the ordinary sense as used in the literature on
liquids and solutions.''

Another issue is that the analogy does not specify
a suitable reference state for proteins. This is also reflected in
the literature on statistical potentials; the construction of a suitable
reference state continues to be an active research topic \cite{pmid14739325,pmid17075131,pmid17351015,pmid17335003,pmid19127590,rykunov2010new}.
In the next section, we discuss a second, more recent justification
that is based on probabilistic reasoning.

\subsection*{PMFs from likelihoods}

Baker and co-workers\emph{ }\cite{pmid9149153} justified PMFs from a
Bayesian point of view and used these insights in the construction of
the coarse grained ROSETTA energy function; Samudrala and Moult used
similar reasoning for the RAPDF potential \cite{pmid9480776}.  According
to Bayesian probability calculus, the conditional probability $P(X\mid
A)$ of a structure $X$, given the amino acid sequence $A$, can be
written as: \[ P\left(X\mid A\right)=\frac{P\left(A\mid
X\right)P\left(X\right)}{P\left(A\right)}\propto P\left(A\mid
X\right)P\left(X\right)\] $P(X\mid A)$ is proportional to the product of
the likelihood $P\left(A\mid X\right)$ times the prior
$P\left(X\right)$. By assuming that the likelihood can be approximated
as a product of pairwise probabilities, and applying Bayes' theorem, the
likelihood can be written as: \begin{equation} P\left(A\mid
X\right)\approx\prod_{i<j}P\left(a_{i},a_{j}\mid
r_{ij}\right)\propto\prod_{i<j}\frac{P\left(r_{ij}\mid
a_{i},a_{j}\right)}{P(r_{ij})}\label{eq_rosettapairs}\end{equation}
where the product runs over all amino acid pairs $a_{i},a_{j}$ (with
$i<j$), and $r_{ij}$ is the distance between amino acids $i$ and $j$.
Obviously, the negative of the logarithm of expression
(\ref{eq_rosettapairs}) has the same functional form as the classic
pairwise distance PMFs, with the denominator playing the role of the
reference state in Eq.\ \ref{eq:classic}.  The merit of this explanation
is the qualitative demonstration that the functional form of a PMF can
be obtained from probabilistic reasoning.  Although this view is
insightful -- it rightfully drew the attention to the application of
Bayesian methods to protein structure prediction -- there is a more
quantitative explanation, which does not rely on the incorrect
assumption of pairwise decomposability
\cite{pearl_pairwise,pmid8609636,ben-naim:3698,pmid9526121}, and leads
to a different, \emph{quantitative} conclusion regarding the nature of
the reference state. This explanation is given in the next section.

\subsection*{A general statistical justification for PMFs}

\begin{figure}[!ht]
\begin{centering}
\includegraphics[clip,width=0.9\columnwidth]{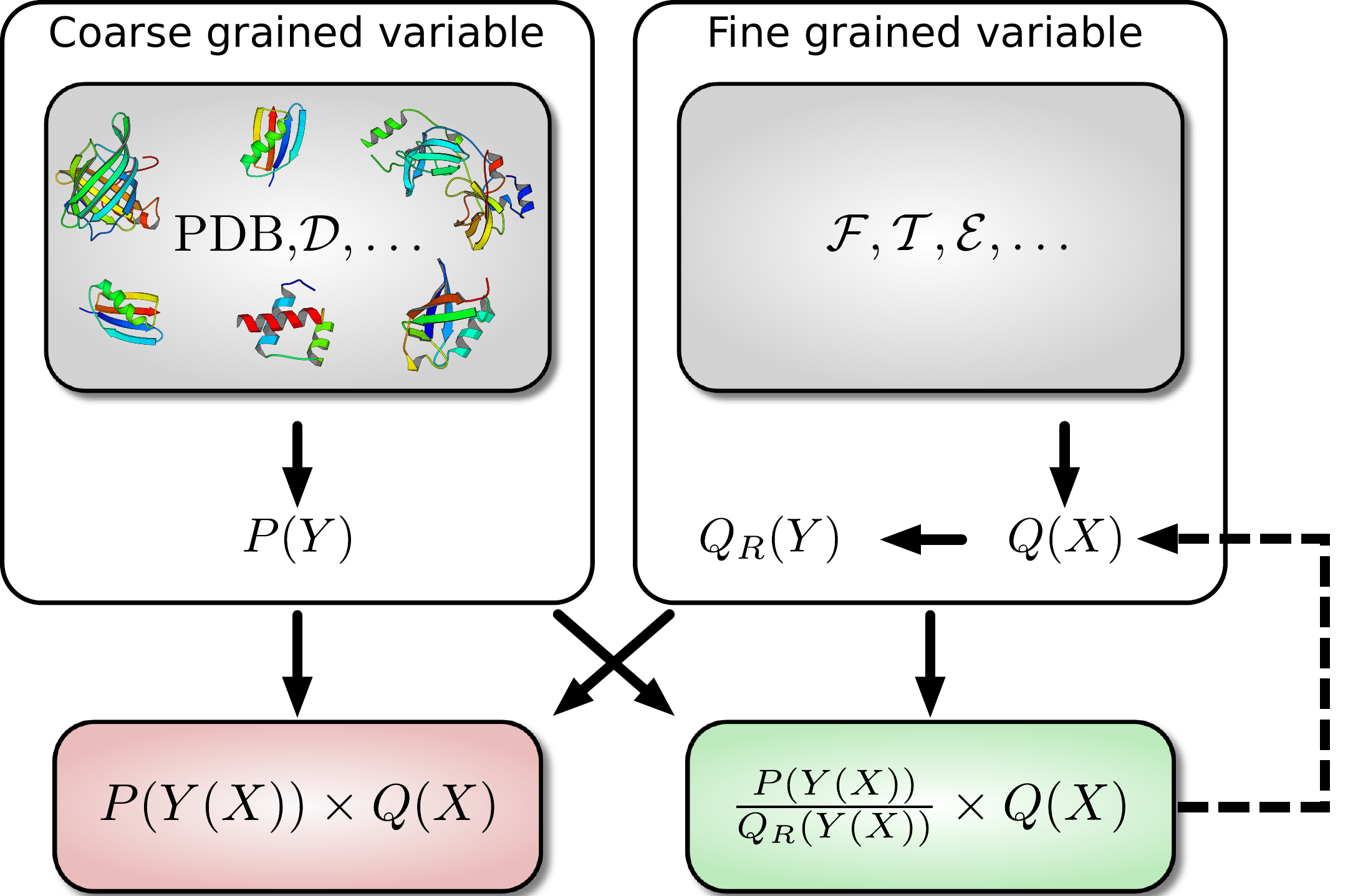} 
\par\end{centering}

\caption{\label{fig:full}\textbf{General statistical justification of
PMFs.} The goal is to combine a distribution $Q(X)$ over a fine grained
variable $X$ (top right), with a probability distribution $P(Y)$ over a
coarse grained variable $Y(X)$ (top left). $Q(X)$ could be, for
example, embodied in a fragment library ($\mathcal{F}$), a probabilistic
model of local structure ($\mathcal{T}$) or an energy function
($\mathcal{E}$); $Y$ could be, for example, the radius of gyration, the
hydrogen bond network, or the set of pairwise distances.  $P(Y)$ usually
reflects the distribution of $Y$ in known protein structures (PDB), but
could also stem from experimental data ($\mathcal{D}$). Sampling from
$Q(X)$ results in a distribution $Q_{R}(Y)$ that differs from $P(Y)$.
Multiplying $P(Y)$ and $Q(X)$ does not result in the desired
distribution for $Y$ either (red box); the correct result requires
dividing out the signal with respect to $Y$ due to $Q(X)$ (green box).
The \emph{reference} distribution $Q_{R}(Y)$ in the denominator
corresponds to the contribution of the reference state in a PMF. If
$Q_{R}(Y)$ is only approximately known, the method can be applied
iteratively (dashed arrow). In that case, one attempts to iteratively
sculpt an energy funnel. The procedure is statistically rigorous
provided $Q(X)$ and $P(Y)$ are proper probability distributions; this is
usually not the case for conventional pairwise distance PMFs.}

\end{figure}

Expressions that resemble PMFs naturally result from the application of
probability theory to solve a fundamental problem that arises in protein
structure prediction: how to improve an imperfect probability
distribution $Q(X)$ over a first variable $X$ using a probability
distribution $P(Y)$ over a second variable $Y$ (see Fig.\ \ref{fig:full},
Fig.\ \ref{fig:simple} and Materials and Methods). We assume that $Y$ is
a deterministic function of $X$; we write $Y(X)$ when necessary. In that
case, $X$ and $Y$ are called \emph{fine} and \emph{coarse grained
variables}, respectively. When $Y$ is a function of $X$, the probability
distribution $Q(X)$ automatically implies a probability distribution
$Q(X,Y(X))$. This distribution has some unusual properties:
$Q(X,Y(X))=Q(X)$; and if $Y'\neq Y(X)$, it follows that $Q(X,Y')=0$.

Typically, $X$ represents \emph{local} features of protein
structure (such as backbone dihedral angles), while $Y$ represents
\emph{nonlocal} features (such as hydrogen bonding, compactness or
pairwise distances).  However, the same reasoning also applies to other
cases; for example, $P(Y)$ could represent information coming from
experimental data, and $Q(X)$ could be embodied in an empirical force
field as used in molecular mechanics
\cite{moult1997comparison,pmid10753811} (see Fig.\ \ref{fig:full}). 

Typically, the distribution $Q(X)$ in itself is not sufficient for
protein structure prediction: it does not consider important nonlocal
features such as hydrogen bonding, compactness or favorable amino acid
interactions. As a result, $Q(X)$ is incorrect with respect to $Y$, and
needs to be supplemented with a probability distribution $P(Y)$ that
provides additional information. By construction, $P(Y)$ is assumed to
be correct (or at least useful).

The above situation arises naturally in protein structure prediction.
For example, $P(Y)$ could be a probability distribution over the radius
of gyration, hydrogen bond geometry or the set of pairwise distances,
and $Q(X)$ could be a fragment library \cite{pmid9149153} or a
probabilistic model of local structure \cite{pmid18579771}. In Fig.\
\ref{fig:simple}, we used the example of a distribution over the radius
of gyration for $P(Y)$ and a fragment library for $Q(X)$. Obviously,
sampling from a fragment library and retaining structures with the
desired nonlocal structure (radius of gyration, hydrogen bonding, etc.)
is in principle possible, but in practice extremely inefficient. 

How can $Q(X)$ be combined with $P(Y)$ in a meaningful way? As mentioned
previously, simply multiplying the two distributions -- resulting in
$P(Y(X))Q(X)$ -- does not lead to the desired result as the two variables are
obviously not independent. The correct solution follows from simple statistical
considerations (see Materials and Methods), and is given by the following
expression: \begin{eqnarray} P(X) & = &
\frac{P\left(Y(X)\right)}{Q_{R}\left(Y(X)\right)}Q(X)\label{eq:ratio}\end{eqnarray}

\noindent We use the notation $P(X)$, as this distribution implies
the desired distribution $P(Y)$ for $Y(X)$. The distribution $Q_{R}(Y)$ in the
denominator is the probability distribution that is implied by $Q(X)$
over the coarse grained variable $Y$.  Conceptually, dividing by
$Q_{R}(Y)$ takes care of the signal in $Q(X)$ with respect to the coarse
grained variable $Y$. The ratio in this expression corresponds to the
probabilistic formulation of a PMF, and $Q_{R}(Y)$ corresponds to the
reference state (see Materials and Methods). 

In practice, $Q(X)$ is typically not evaluated directly, but brought in through
conformational Monte Carlo sampling (see Materials and Methods); often sampling
is based on a fragment library \cite{sippl1992assembly,pmid9149153}, although
other methods are possible, including sampling from a probabilistic model
\cite{pmid17002495,pmid18579771,Zhao:2010:J-Comput-Biol:20583926} or a suitable
energy function \cite{moult1997comparison,pmid10753811}.  The ratio
$P(Y)/Q_{R}(Y)$, which corresponds to the probabilistic formulation of a PMF,
also naturally arises in the Markov chain Monte Carlo (MCMC) procedure (see
Materials and Methods). An important insight is that, in this case, the
conformational sampling method uniquely defines the reference state. Thus, in
the case of a fragment library, the reference distribution $Q_{R}(Y)$ is the
probability distribution over $Y$ that is obtained by sampling conformations
solely using the fragment library.

As the method we have introduced here invariably relies on the ratio of two
probability distributions -- one regarding protein structure and the
other regarding a well-defined reference state -- we refer to it as the
\emph{reference ratio method}. In the next section, we show that the standard
pairwise distance PMFs can be seen as an approximation of the reference
ratio method. 

\subsection*{Pairwise distance PMFs explained}

In this section, we apply the reference ratio method to the
standard, pairwise distance case. In the classic PMF approach, one
considers the vector of pairwise distances $R$ between the amino acids.
In this case, it is usually assumed that we can write\begin{equation}
P(R\mid A)\propto\prod_{i<j}P(r_{ij}\mid
a_{i},a_{j})\label{eq:pairwise}\end{equation}

\noindent where the product runs over all amino acid pairs $a_{i},a_{j}$
(with $i<j$), and $r_{ij}$ is their matching distance. Clearly,
the assumption that the joint probability can be written as a product
of pairwise probabilities is not justified \cite{pearl_pairwise,pmid8609636,ben-naim:3698},
but in practice this assumption often provides useful results \cite{rykunov2010new}.
In order to obtain protein-like conformations, $P(R\mid A)$ needs
to be combined with an appropriate probability distribution $Q(X\mid A)$
that addresses the local features of the polypeptide chain. Applying
Eq.\ \ref{eq:ratio} to this case results in the following expression:\[
P(X\mid A)\propto\frac{\prod_{i<j}P(r_{ij}\mid a_{i},a_{j})}{\prod_{i<j}Q_{R}(r_{ij}\mid a_{i},a_{j})}Q(X\mid A)\]

\noindent where the denominator $Q_{R}(\cdot)$ is the probability
distribution over the pairwise distances as induced by the distribution
$Q(X\mid A)$. The ratio in this expression corresponds to the probabilistic
expression of a PMF. The reference state is thus determined by $Q(X\mid A)$:
it reflects the probability of generating a set of pairwise distances
using local structure information alone. Obviously, as $Q(X\mid A)$
is conditional upon the amino acid sequence $A$, the reference state
becomes sequence dependent as well.

We again emphasize that the assumption of pairwise decomposability in
Eq.\ \ref{eq:pairwise} is incorrect
\cite{pearl_pairwise,pmid8609636,ben-naim:3698,pmid9526121}. Therefore,
the application of the reference ratio method results in a useful approximation, at
best.  As a result, the optimal definition of the reference state also
needs to compensate for the errors implied by the invalid assumption. As
is it well established that distance dependent PMFs perform well with a
suitable definition of the reference state
\cite{pmid14739325,pmid17075131,pmid17351015,pmid17335003,rykunov2010new},
and the incorrect pairwise decomposability assumption impairs a rigorous
statistical analysis, we do not discuss this type of PMFs further.
Indeed, for pairwise distance PMFs, the main challenge lies in
developing better probabilistic models of sets of pairwise distances
\cite{pmid19153168}. 

The pairwise distance PMFs currently used in protein structure
prediction are thus not statistically rigorous, because they do not make
use of a proper joint probability distribution over the pairwise
distances, which are strongly intercorrelated due to the connectivity of
molecules. A rigorous application of the reference ratio method would require
the construction of a proper joint probability distribution over
pairwise distances. This is certainly possible in principle, but
currently, as far as we know, a challenging open problem and
beyond the scope of this article.  However, we have clarified that the
idea of using a reference state is correct and valid, and that
this state has a very precise definition. Therefore, in the next two
sections, we show instead how statistically valid quantities, similar
to PMFs, can be obtained for very different coarse grained variables.

\subsection*{A generalized PMF: radius of gyration}

As a first application of the reference ratio method, we consider the task of sampling protein conformations
with a given probability distribution $P(r_{g})$ for the radius of
gyration $r_{g}$. For $P(r_{g})$, we chose a Gaussian distribution
with mean $\mu=22\,\textrm{\AA}$ and standard deviation
$\sigma=2\,\textrm{\AA}$. This choice is completely arbitrary; it simply serves
to illustrate that the reference ratio method allows imposing an exact
probability distribution over a certain feature of interest.
Applying Eq.\ \ref{eq:ratio} results in:\begin{equation}
P(X \mid A)=\frac{P(r_{g}(X))}{Q_{R}(r_{g}(X)\mid A)}Q(X\mid A)\label{eq:rg}\end{equation}

\noindent For $Q(X\mid A)$, we used TorusDBN -- a graphical model
that allows sampling of plausible backbone angles \cite{pmid18579771}
-- and sampled conditional on the amino acid sequence $A$ of ubiquitin
(see Materials and Methods). $Q_{R}(r_{g}\mid A)$ is the probability
distribution of the radius of gyration for structures sampled solely
from TorusDBN, which was determined using generalized multihistogram
MCMC sampling (see Materials and Methods).

\begin{figure}[!ht]
\begin{centering}
\includegraphics[width=0.9\columnwidth]{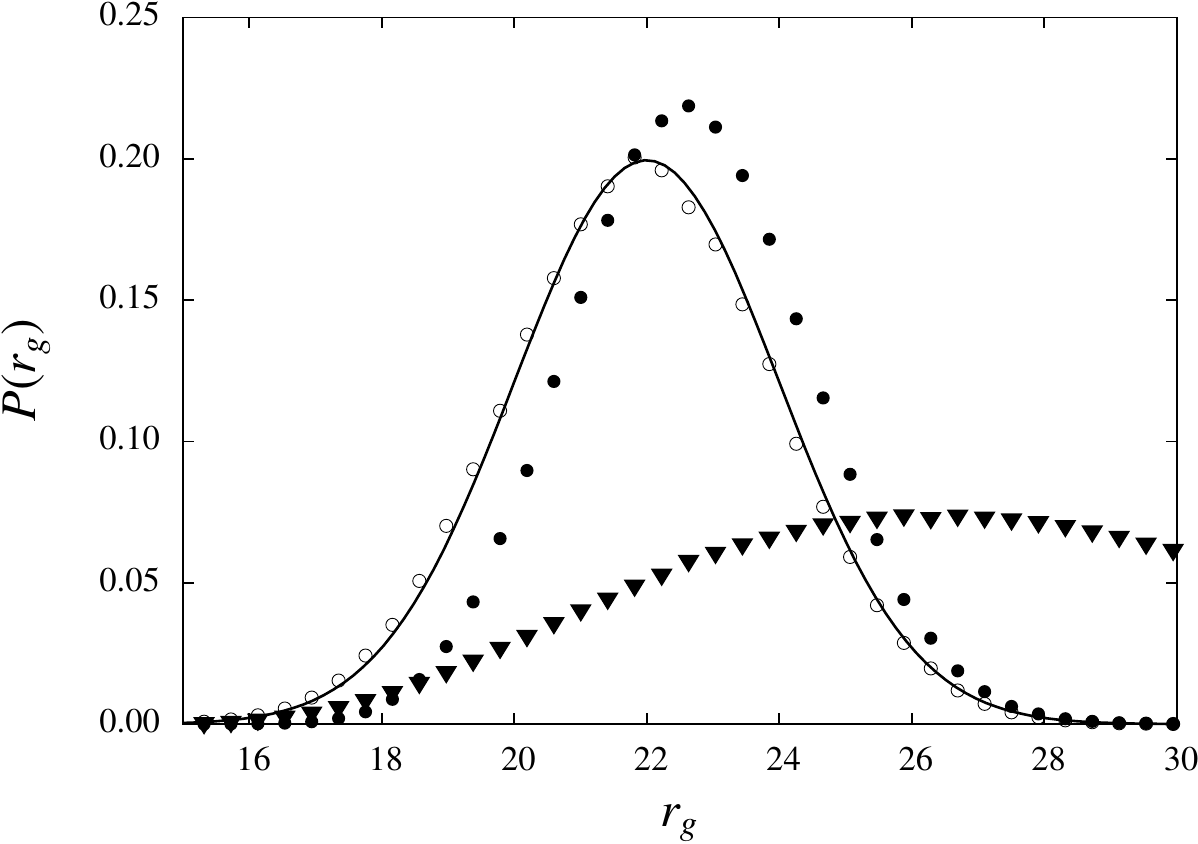}
\par\end{centering}

\caption{\label{fig:rg_plot}\textbf{A PMF based on the radius of gyration.}
The goal is to adapt a distribution $Q(X\mid A)$ -- which allows
sampling of local structures -- such that a given target distribution
$P(r_{g})$ is obtained. For $A$, we used the amino acid sequence of
ubiquitin.  Sampling from $Q(X\mid A)$ alone results
in a distribution with an average $r_{g}$ of about 27 $\textrm{\AA}$
(triangles). Sampling using the correct expression (open circles),
given by Eq.\ \ref{eq:rg}, results in a distribution that coincides
with the target distribution (solid line). Not taking the reference
state into account results in a significant shift towards higher $r_{g}$
(black circles). }

\end{figure}

In Fig.\ \ref{fig:rg_plot}, we contrast sampling from Eq.\ \ref{eq:rg}
with sampling from $P(r_{g}(X))Q(X\mid A)$. In the latter case, the
reference state is not properly taken into account, which results
in a significant shift towards higher radii of gyration. In contrast,
the distribution of $r_{g}$ for the correct distribution $P(X)$,
given by Eq.\ \ref{eq:rg}, is indistinguishable from the target distribution.
This qualitative result is confirmed by the Kullback-Leibler divergence
\cite{kullback1951information} -- a natural distance measure for
probability distributions expressed in bits -- between the target
distribution and the resulting marginal distributions of $r_{g}$.
Adding $Q_{R}(r_{g}(X)\mid A)$ to the denominator diminishes the
distance from 0.08 to 0.001 bits. For this particular PMF, the effect
of using the correct reference state is significant, but relatively
modest; in the next section, we discuss an application where its effect
is much more pronounced.

\subsection*{Iterative optimization of PMFs: hydrogen bonding}

Here, we demonstrate that PMFs can be optimized iteratively, which is
particularly useful if the reference probability distribution
$Q_{R}(Y\mid A)$ is difficult to estimate. We illustrate the method with
a target distribution that models the hydrogen bonding network using a
multinomial distribution.

We describe the hydrogen bonding network ($H$) with eight integers
(for details, see Materials and Methods). Three integers $(n_{\alpha},n_{\beta},n_{c})$
represent the number of residues that do not partake in hydrogen bonds
in $\alpha$-helices, $\beta$-sheets and coils, respectively. The
five remaining integers $(n_{\alpha\alpha},n_{\beta\beta},n_{cc},n_{\alpha c},n_{\beta c})$
represent the number of hydrogen bonds within $\alpha$-helices, within
$\beta$-strands, within coils, between $\alpha$-helices and coils,
and between $\beta$-strands and coils, respectively.

As target distribution $P(H)$ over these eight integers, we chose a
multinomial distribution whose parameters were derived from the native
structure of protein G (see Materials and Methods). $P(H)$ provides
information, regarding protein G, on the number of hydrogen bonds and the
secondary structure elements involved, but does not specify \emph{where}
the hydrogen bonds or secondary elements occur. As in the previous
section, we use TorusDBN as the sampling distribution $Q(X\mid A)$; we
sample backbone angles conditional on the amino acid sequence $A$ of
protein G. Native secondary structure information was \emph{not} used in
sampling from TorusDBN.

The reference distribution $Q_{R}(H\mid A)$, due to TorusDBN, is very
difficult to estimate correctly for several reasons: its shape is
unknown and presumably complex; its dimensionality is high; and the data
is very sparse with respect to $\beta$-sheet content. Therefore,
$Q_{R}(H\mid A)$ can only be approximated, which results in a suboptimal
PMF. A key insight is that one can apply the method iteratively until a
satisfactory PMF is obtained (see Fig.\ \ref{fig:full}, dashed line). In
each iteration, the (complex) reference distribution is approximated
using a simple probability distribution;  we illustrate the method by
using a multinomial distribution, whose parameters are estimated by
maximum likelihood estimation in each iteration, using the conformations
generated in the previous iteration. In the first iteration, we simply
set the reference distribution equal to the uniform distribution.

Formally, the procedure works as follows. In iteration $i+1$, the
distribution $P_{i}(H \mid A)$ is improved using the samples generated in
iteration $i$: \begin{equation}
P_{i+1}(X\mid A)=\frac{P(H(X))}{P_{R,i}(H(X) \mid A)}P_{i}(X \mid A)\label{eq:hbond_iterative}\end{equation}

\noindent where $P_{R,i}(H \mid A)$ is the reference distribution
estimated from the samples generated in the $i$-th iteration,
$P_{0}(X)=Q(X\mid A)$ stems from TorusDBN, and $P_{R,0}(H \mid A)$ is
the uniform distribution. After each iteration, the set of samples is
enriched in hydrogen bonds, and the reference distribution $P_{R,i}(H
\mid A)$ can be progressively estimated more precisely. Note that in the
first iteration, we simply use the product of the target and the
sampling distribution; no reference state is involved.

\begin{figure}[!ht]
\begin{centering}
\includegraphics[width=0.9\columnwidth]{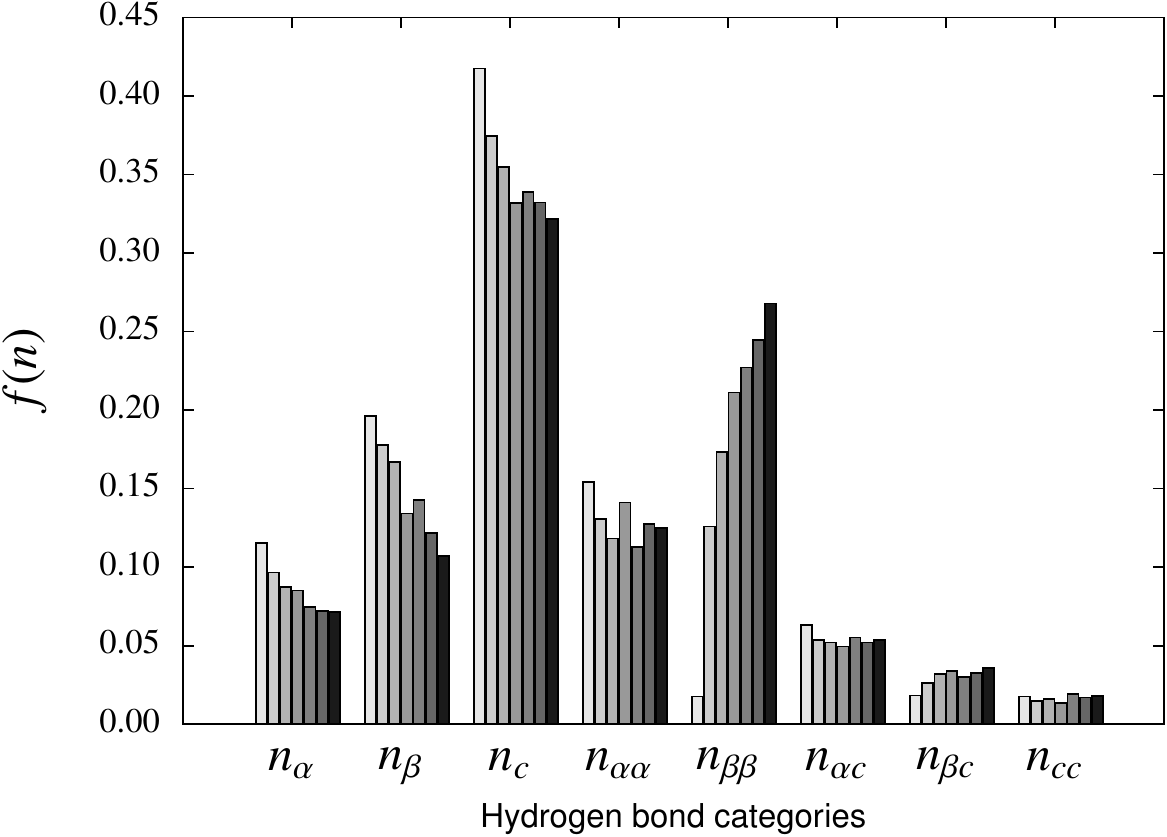}
\par\end{centering}

\caption{\label{fig:hbond_counts}\textbf{Iterative estimation of a PMF.} For each
of the eight hydrogen bond categories (see text), the black bar to
the right denotes the fraction of occurrence $f(n)$ in the native
structure of protein G. The gray bars denote the fractions of the
eight categories in samples from each iteration; the first iteration
is shown to the left in light gray. In the last iteration (iteration
6; dark gray bars, right) the values are very close to the native
values for all eight categories. Note that hydrogen bonds between
$\beta$-strands are nearly absent in the first iteration (category
$n_{\beta\beta}$). }

\end{figure}

\begin{figure*}[!ht]
\begin{centering}
\includegraphics[width=0.7\paperwidth]{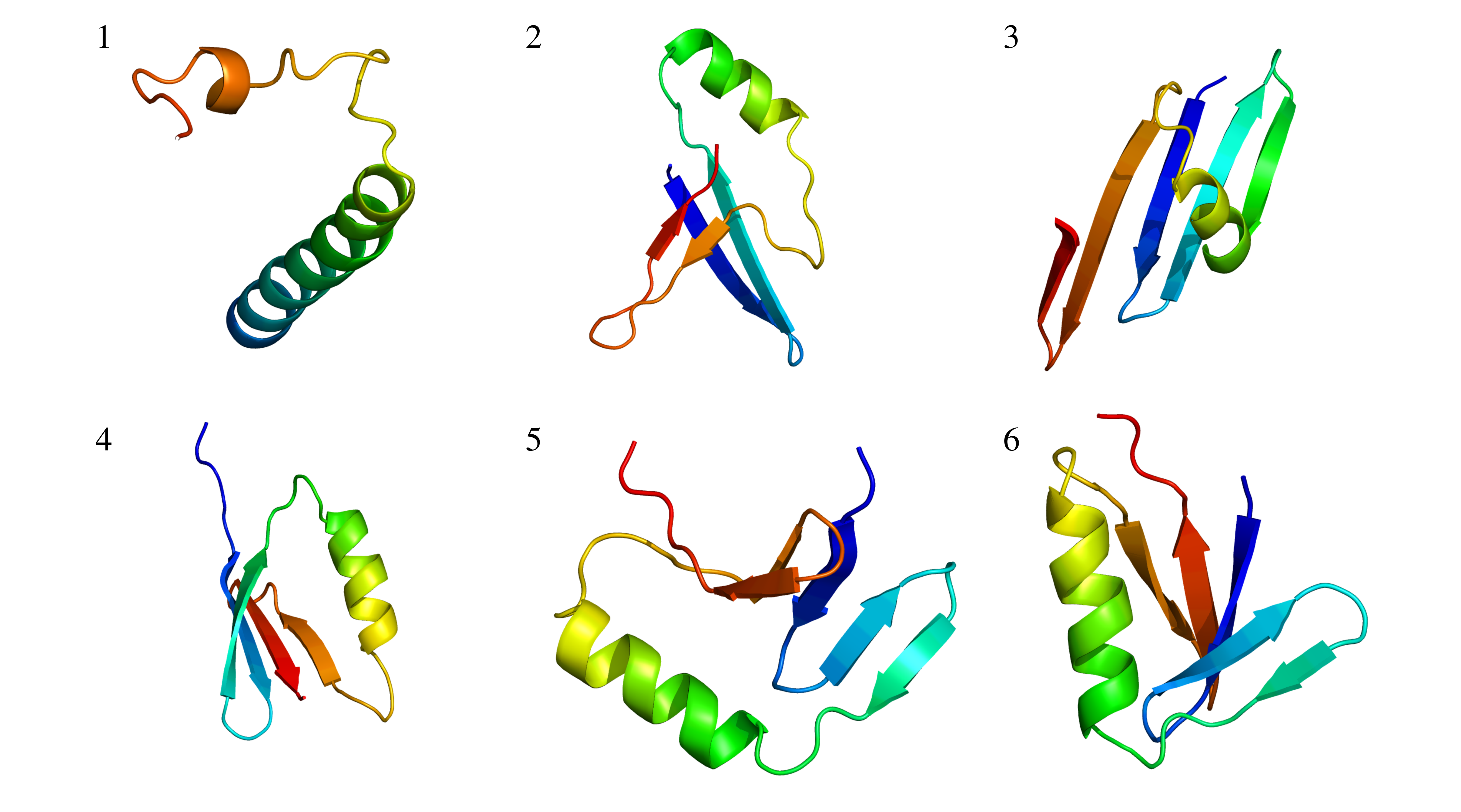}
\par\end{centering}

\centering{}\caption{\label{fig:hbond_structures}\textbf{Highest
probability structures for each iteration.} The structures with highest
probability out of 50,000 samples for all six iterations (indicated by a
number) are shown as cartoon representations. The N-terminus is shown in
blue.  The figure was made using PyMOL \cite{DeLano2002}. }

\end{figure*}

Fig.\ \ref{fig:hbond_counts} shows the evolution of the fractions
versus the iteration number for the eight hydrogen bond categories;
the structures with minimum energy for all six iterations are shown
in Fig.\ \ref{fig:hbond_structures}. In the first iteration, the structure
with minimum energy (highest probability) consists of a single $\alpha$-helix;
$\beta$-sheets are entirely absent (see Fig.\ \ref{fig:hbond_structures},
structure 1). Already in the second iteration, $\beta$-strands start to pair,
and in the third and higher iterations complete sheets are readily
formed. The iterative optimization of the PMF quickly leads to a dramatic
enrichment in $\beta$-sheet structures, as desired, and the fractions
of the eight categories become very close to the native values (Fig.\
\ref{fig:hbond_counts}). 

\noindent %

\subsection*{Conclusions}

The strengths and weaknesses of PMFs can be rigorously explained based on
simple probabilistic considerations, which leads to some surprising new
insights of direct practical relevance.  First, we have made clear that
PMFs naturally arise when two probability distributions need to be
combined in a meaningful way. One of these distributions typically
addresses local structure, and its contribution often arises from
conformational sampling. Each conformational sampling method thus
requires its own reference state and corresponding reference
distribution; this is likely the main reason behind the large number of
different reference states reported in the literature
\cite{pmid14739325,pmid17075131,pmid17351015,pmid17335003,pmid19127590,rykunov2010new}.
If the sampling method is conditional upon the amino acid sequence, the
reference state necessarily also depends on the amino acid sequence.

Second, conventional applications of pairwise distance PMFs usually lack two
necessary features to make them fully rigorous: the use of a proper
probability distribution over pairwise distances in proteins for $P(Y
\mid A)$, and the recognition that the reference state is rigorously
defined by the conformational sampling scheme used, that is, $Q(X \mid
A)$. Usually, the reference state is derived from 
external physical considerations \cite{rooman1995database, pmid12381853}.

Third, PMFs are not tied to pairwise distances, but generalize to any
coarse grained variable. Attempts to develop similar quantities that,
for example, consider solvent exposure \cite{bowie1991method,
liithy1992assessment}, relative side chain orientations
\cite{pmid15093838}, backbone dihedral angles
\cite{rooman1991prediction, kocher1994factors} or hydrogen bonds
\cite{pmid9079391} are thus, in principle, entirely justified. Hence, our
probabilistic interpretation opens up a wide range of possibilities for
advanced, well-justified energy functions based on sound probabilistic
reasoning; the main challenge is to develop proper probabilistic models
of the features of interest and the estimation of their parameters
\cite{pmid10336385,pmid19153168}.  Strikingly, the example applications
involving radius of gyration and hydrogen bonding that we presented in
this article \emph{are} statistically valid and rigorous, in contrast to
the traditional pairwise distance PMFs.

Finally, our results reveal a straightforward way to optimize PMFs.
Often, it is difficult to estimate the probability distribution that
describes the reference state. In that case, one can start with an
approximate PMF, and apply the method iteratively. In each iteration,
a new reference state is estimated, with a matching probability distribution.
In that way, one iteratively attempts to sculpt an energy funnel \cite{bryngelson1987spin,leopold1992protein,dill1997levinthal,pmid12926006,fain2003funnel,pmid19291741}.
We illustrated this approach with a probabilistic model of the hydrogen
bond network. Although iterative application of the inverse Boltzmann
formula has been described before \cite{thomas1996iterative,pmid12926006,huang2006iterative,pmid19291741},
its theoretical justification, optimal definition of the reference
state and scope remained unclear.

As the traditional pairwise distance PMFs used in protein structure
prediction arise from the imperfect application of a statistically valid
and rigorous procedure with a much wider scope, we consider it highly
desirable that the name {}``potential of mean force'' should be reserved
for true, physically valid quantities \cite{ben-naim:3698}. Because the
statistical quantities we discussed invariably rely on the use of a
ratio of two probability distributions, one concerning protein structure
and the other concerning the (now well defined) reference state, we
suggest the name {}``reference ratio distribution'' deriving from the
application of the {}``reference ratio method''. 

Pairwise distance PMFs, as used in protein structure prediction, are not
physically justified potentials of mean force or free energies
\cite{moult1997comparison,ben-naim:3698} and the reference state does
not depend on external physical considerations; the same is of course
true for our generalization. However, these PMFs are approximations of
statistically valid and rigorous quantities, and these quantities can be
generalized beyond pairwise distances to other aspects of protein
structure. The fact that these quantities are not potentials of mean
force or free energies is of no consequence for their statistical rigor
or practical importance -- both of which are considerable. Our results
thus vindicate, formalize and generalize Sippl's original and seminal
idea \cite{pmid2359125}. After about twenty years of controversy, PMFs -- or
rather the statistical quantities that we have introduced in this
article -- are ready for new challenges.

\section*{Materials and methods}

\subsection*{Outline of the problem}

We consider a joint probability distribution $Q(X,Y)$ and a probability
distribution $P(Y)$ over two variables of interest, $X$ and $Y$, where
$Y$ is a deterministic function of $X$;  we write $Y(X)$ when relevant.
Note that because $Y$ is a function of $X$, it follows that
$Q(X)=Q(X,Y(X))$; and if $Y'\neq Y(X)$, then $Q(X,Y')=0$. 

We assume that $P(Y)$ is a meaningful and informative distribution for
$Y$. Next, we note that $Q(X,Y)$ implies a matching marginal probability
distribution $Q_{R}(Y)$ (where the subscript $R$ refers to the fact
that $Q_{R}(Y)$ corresponds to the reference state, as we will show
below):\[ Q_{R}(Y)=\intop Q(X,Y)\textrm{d}X\] We consider the case
where $Q_{R}(Y)$ differs substantially from $P(Y)$; hence, $Q_{R}(Y)$
can be considered as incorrect. On the other hand, we also assume that
the conditional distribution $Q(X\mid Y)$ is indeed meaningful and
informative (see next section). This distribution is given by: 
\begin{equation}
Q(X\mid Y)=
\begin{cases} 
0 & \text{if $Y\neq Y(X)$} \\
\frac{Q(X)}{\intop Q(X')\delta(Y(X')-Y)\textrm{d}X'} & \text{if
$Y=Y(X)$}
\end{cases}
\end{equation}
\noindent where $\delta(\cdot)$ is the delta function. The question is now how to combine the two distributions $P(Y)$ and
$Q(X)$ -- each of which provide useful information on $X$ and $Y$ -- in
a meaningful way. Before we provide the solution, we illustrate how this
problem naturally arises in protein structure prediction.

\subsection*{Application to protein structure}

In protein structure prediction, $Q(X,Y)$ is often embodied in a
fragment library; in that case, $X$ is a set of atomic coordinates
obtained from assembling a set of polypeptide fragments. Of course,
$Q(X,Y)$ could also arise from a probabilistic model, a pool of known
protein structures, or any other conformational sampling method. The
variable $Y$ could, for example, be the radius of gyration, the hydrogen
bond network or the set of pairwise distances. If $Y$ is a deterministic
function of $X$, the two variables are called \emph{coarse grained}
and \emph{fine grained} variables, respectively. For example, sampling
a set of dihedral angles for the protein backbone uniquely defines
the hydrogen bond geometry between any of the backbone atoms.

Above, we assumed that $Q(X\mid Y)$ is a meaningful distribution.
This is often a reasonable assumption; fragment libraries, for example,
originate from real protein structures, and conditioning on protein-like
compactness or hydrogen bonding will thus result in a meaningful distribution.
Of course, sampling solely from $Q(X,Y)$ is not an efficient strategy
to obtain hydrogen bonded or compact conformations, as they will be
exceedingly rare. We now provide the solution of the problem outlined
in the previous section, and discuss its relevance to the construction
of PMFs.

\subsection*{Solution for a proper joint distribution}

A first step on the way to the solution is to note that the product
rule of probability theory allows us to write: 

\[
P(X,Y)=P(Y)P(X\mid Y)\]

\noindent As only $P(Y)$ is given, we need to make a reasonable choice
for $P(X\mid Y)$. We assume, as discussed before, that $Q(X\mid Y)$
is a meaningful choice, which leads to:\[
P(X,Y)=P(Y)Q(X\mid Y)\]

\noindent In the next step, we apply the product formula of probability
theory to the second factor $Q(X\mid Y)$, and obtain:\begin{equation}
P(X,Y)=P(Y)\frac{Q(X,Y)}{Q_{R}(Y)}\label{eq:BH}\end{equation}

\noindent The distribution $P(X,Y)$ has the correct marginal distribution
$P(Y)$.

In the next two sections, we discuss how this straightforward result can be
used to great advantage for understanding and generalizing PMFs. First, we show
that the joint distribution specified by Eq.\ \ref{eq:BH} can be reduced to a
surprisingly simple functional form.  Second, we discuss how this result can be
used in MCMC sampling. In both cases, expressions that correspond to a PMF
arise naturally.

\subsection*{PMFs from combining distributions}

\noindent Using the product rule of probability theory, Eq.\ \ref{eq:BH}
can be written as:\[ P(X,Y)=P(Y)\frac{Q(Y\mid X)Q(X)}{Q_{R}(Y)}\]
Because the coarse grained variable $Y$ is a deterministic function of
the fine grained variable $X$, $Q(Y\mid X)$ is the delta function:

\begin{equation}
P(X,Y)=P(Y)\frac{\delta\left(Y-Y(X)\right)Q(X)}{Q_{R}(Y)}\end{equation}

\noindent Finally, we integrate out the, now redundant, coarse grained
variable $Y$ from the expression:\begin{eqnarray*}
P(X) & = & \int P(X,Y)\textrm{d}Y\\
 & = & \int P(Y)\frac{\delta\left(Y-Y(X)\right)Q(X)}{Q_{R}(Y)}\textrm{d}Y\\
 & = & \frac{P(Y(X))}{Q_{R}(Y(X))}Q(X)\end{eqnarray*}
and obtain our central result (Eq.\ \ref{eq:ratio}). Sampling from
$P(X)$ will result in the desired marginal probability distribution
$P(Y)$. The influence of the fine grained distribution $Q(X,Y)$
is apparent in the fact that $P(X\mid Y)$ is equal to $Q(X\mid Y)$.
The ratio in this expression corresponds to the usual probabilistic
formulation of a PMF; the distribution $Q_{R}(Y)$ corresponds to
the reference state. In the next section, we show that PMFs also naturally
arise when $P(Y)$ and $Q(X,Y)$ are used together in Metropolis-Hastings
sampling.

\subsection*{PMFs from Metropolis-Hastings sampling}

Here, we show that Metropolis-Hastings sampling from the distribution
specified by Eq.\ \ref{eq:BH}, using $Q(X,Y)$ as a proposal distribution,
naturally results in expressions that are equivalent to PMFs. The
derivation is also valid if the proposal distribution depends on the
previous state, provided $Q(X,Y)$ satisfies the detailed balance condition.

According to the standard Metropolis-Hastings method
\cite{gilks1996markov}, one can sample from a probability distribution
$\Pi(X,Y)$ by generating a Markov chain where each state $X',Y'$ depends
only on the previous state $X,Y$. The new state $X',Y'$ is generated
using a proposal distribution $\pi(Y',X'\mid Y,X)$, which includes
$\pi(X',Y'\mid X,Y)=\pi(X',Y')$ as a special case. According to the
Metropolis-Hastings method, the proposal $X',Y'$ is accepted with a
probability $\alpha$:\[ \alpha(X',Y'\mid X,Y)=\min(1,p),\]

\begin{equation}
p=\frac{\Pi(X',Y')}{\Pi(X,Y)}\times\frac{\pi(X,Y\mid X',Y')}{\pi(X',Y'\mid X,Y)}\label{eq:MetHas}\end{equation}

\noindent where $Y,X$ is the starting state, and $Y',X'$ is the next proposed
state. We assume that the proposal distribution $\pi(X',Y'\mid X,Y)$ 
satisfies the detailed balance condition:\[ \pi(X',Y'\mid
X,Y)\pi(X,Y)=\pi(X,Y\mid X',Y')\pi(X',Y')\]

\noindent As a result, we can always write Eq.\ \ref{eq:MetHas} as:\[
\frac{\Pi(X',Y')}{\Pi(X,Y)}\times\frac{\pi(X,Y)}{\pi(X',Y')}\]

\noindent The Metropolis-Hastings expression (Eq.\ \ref{eq:MetHas}), applied to the distribution
specified by Eq.\ \ref{eq:BH} and using $Q(X',Y')$ or $Q(X',Y'\mid X,Y)$
as the proposal distribution, results in: 

\[
\frac{P(Y')Q_{R}(Y)Q(X',Y')}{P(Y)Q_{R}(Y')Q(X,Y)}\times\frac{Q(X,Y)}{Q(X',Y')}\]

\noindent which reduces to:\begin{equation}
\frac{P(Y')}{P(Y)}\times\frac{Q_{R}(Y)}{Q_{R}(Y')}\label{eq:MHratio}\end{equation}

\noindent Hence, we see that the Metropolis-Hastings method requires the
evaluation of ratios of the form $P(Y)/Q_{R}(Y)$ when $Q(X',Y')$ or
$Q(X',Y'\mid X,Y)$ is used as the proposal distribution; these ratios
correspond to the usual probabilistic formulation of a PMF.  Finally,
when $Y$ is a deterministic function of $X$, the proposal distribution
reduces to $Q(X')$ or $Q(X'\mid X)$, and Eq.\ \ref{eq:MHratio} becomes:\[
\frac{P(Y(X'))}{P(Y(X))}\times\frac{Q_{R}(Y(X))}{Q_{R}(Y(X'))}\]

\subsection*{Application to radius of gyration and hydrogen bonding}

Conformational sampling from a suitable $Q(X \mid A)$ was done using TorusDBN
\cite{pmid18579771} as implemented in Phaistos \cite{PHAISTOS_LASR09};
backbone angles ($\phi,\psi$ and $\omega$) were sampled conditional
on the amino acid sequence. We used standard fixed bond lengths and
bond angles in constructing the backbone coordinates from the angles,
and represented all side chains (except glycine and alanine) with
one dummy atom with a fixed position \cite{PHAISTOS_LASR09}.

For the radius of gyration application, we first determined $Q_{R}(r_{g}\mid A)$
using the multi-canonical MCMC method to find the sampling weights
$w(r_{g})$ that yield a flat histogram \cite{jfb02}. Sampling from
the resulting joint distribution (Eq.\ \ref{eq:rg}) was done using
the same method. In both cases, we used 50 million iterations; the
$r_{g}$ bin size was 0.08 $\textrm{\AA}$. Sampling from TorusDBN
was done conditional on the amino acid sequence $A$ of ubiquitin
(76 residues, PDB code 1UBQ).

For the hydrogen bond application, sampling from the PMFs was done in
the $1/k$-ensemble \cite{hesselbo1995monte}, using the
Metropolis-Hastings algorithm and the generalized multihistogram method
for updating the weights \cite{jfb02}. In each iteration $i$, 50,000
samples (out of 50 million Metropolis-Hastings steps) were generated,
and the parameters of the multinomial distribution $Q_{R,i}(H)$ were
subsequently obtained using maximum likelihood estimation. Hydrogen
bonds were defined as follows: the $N,O$ distance is below 3.5
$\textrm{\AA}$, and the angles formed by $O,H,N$ and $C,O,H$ are both
greater than 100$^\circ$.  Each carbonyl group was assumed to be
involved in at most one hydrogen bond; in case of multiple hydrogen bond
partners, the one with the lowest $H,O$ distance was selected. Each
residue was assigned to one of the eight possible hydrogen bond
categories
$(n_{\alpha},n_{\beta},n_{c},n_{\alpha\alpha},n_{\beta\beta},n_{cc},n_{\alpha
c},n_{\beta c})$ based on the presence of hydrogen bonding at its
carbonyl group and the secondary structure assignments (for both bond
partners) by TorusDBN.  The target distribution -- the multinomial
distribution $P(H)$ used in Eq.\ \ref{eq:hbond_iterative} -- was obtained
by maximum likelihood estimation using the number of hydrogen bonds, for
all eight categories, in the native structure of protein G (56 residues,
PDB code 2GB1). Sampling from TorusDBN was done conditional on the amino
acid sequence of protein G; native secondary structure information was
\emph{not} used.

\section*{Author contributions}
T.H., M.B. and M.P. are joint first authors. T.H. and J.F.B. developed
the theory. M.B. and M.P. performed the simulations. The remaining
authors contributed new tools. T.H. wrote the paper.

\section*{Acknowledgments}

We acknowledge funding by the \emph{Danish Program Commission on Nanoscience,
Biotechnology and IT }(NaBiIT, project: {}``Simulating proteins on
a millisecond time-scale'', 2106-06-0009), the \emph{Danish Research
Council for Technology and Production Sciences }(FTP, project: {}``Protein
structure ensembles from mathematical models'', 274-09-0184) and
the \emph{Danish Council for Independent Research} (FNU, project:
{}``A Bayesian approach to protein structure determination'', 272-08-0315).


\newpage{}

\section*{Figure legends}


\end{document}